\documentstyle[12pt,amsfonts]{article}

\def\hybrid{\topmargin 0pt      \oddsidemargin 0pt
        \headheight 0pt \headsep 0pt
        \textwidth 6.25in       
        \textheight 9.5in       
        \marginparwidth .875in
        \parskip 5pt plus 1pt   \jot = 1.5ex}
\catcode`\@=11
\def\marginnote#1{}
\newcount\hour
\newcount\minute
\newtoks\amorpm
\hour=\time\divide\hour by60
\minute=\time{\multiply\hour by60 \global\advance\minute
by-\hour}\edef\standardtime{{\ifnum\hour<12
\global\amorpm={am}%
        \else\global\amorpm={pm}\advance\hour by-12 \fi
        \ifnum\hour=0 \hour=12 \fi
        \number\hour:\ifnum\minute<10
0\fi\number\minute\the\amorpm}}
\edef\militarytime{\number\hour:\ifnum\minute<10
0\fi\number\minute}
 
\def\draftlabel#1{{\@bsphack\if@filesw {\let\thepage\relax
   \xdef\@gtempa{\write\@auxout{\string
      \newlabel{#1}{{\@currentlabel}{\thepage}}}}}\@gtempa
   \if@nobreak \ifvmode\nobreak\fi\fi\fi\@esphack}
        \gdef\@eqnlabel{#1}}
\def\@eqnlabel{}
\def\@vacuum{}
\def\draftmarginnote#1{\marginpar{\raggedright\scriptsize\tt#1}}
\def\draft{\oddsidemargin 0pt 
        \def\@oddfoot{\sl preliminary draft \hfil
        \rm\thepage\hfil\sl\today\quad\militarytime}
        \let\@evenfoot\@oddfoot \overfullrule 3pt
        \let\label=\draftlabel
        \let\marginnote=\draftmarginnote
 
\def\@eqnnum{(\theequation)\rlap{\kern\marginparsep\tt\@eqnlabel}%
\global\let\@eqnlabel\@vacuum}  }
 
\def\numberbysection{\@addtoreset{equation}{section}
        \def\theequation{\thesection.\arabic{equation}}}

\def\underline#1{\relax\ifmmode\@@underline#1\else
 $\@@underline{\hbox{#1}}$\relax\fi}

\def\eqnarray{\stepcounter{equation}\let\@currentlabel=\theequation
\global\@eqnswtrue
\global\@eqcnt\z@\tabskip\@centering\let\\=\@eqncr
$$\halign to \displaywidth\bgroup\@eqnsel\hskip\@centering
  $\displaystyle\tabskip\z@{##}$&\global\@eqcnt\@ne 
  \hfil$\displaystyle{\hbox{}##\hbox{}}$\hfil
  &\global\@eqcnt\tw@ $\displaystyle\tabskip\z@
  {##}$\hfil\tabskip\@centering&\llap{##}\tabskip\z@\cr}
\def\lefteqn#1{\hbox to 2em{$\displaystyle #1$\hss}}
\mathchardef\by="202
\def\cn{\mathop{\rm cn}\nolimits}
\def\dn{\mathop{\rm dn}\nolimits}
\def\sn{\mathop{\rm sn}\nolimits}
\makeatletter
\numberbysection
\hybrid

\begin{document}

\begin{titlepage}
\nopagebreak
\begin{flushright}
LPTENS-96/61\\
solv-int/9611003
\\
October 1996
\end{flushright}
\vskip 5cm
\begin{center}
{\large\bf
SOME EXPLICIT SOLUTIONS OF \\[0.5cm] THE LAM\'E AND BOURLET TYPE EQUATIONS}
\vglue 1  true cm
{\bf Alexander V. Razumov}\\
{\footnotesize Institute for High Energy Physics,
142284, Protvino, Moscow region, Russia\footnote{E-mail:
razumov@mx.ihep.su}}\\  
and\\
{\bf Mikhail V. Saveliev}\footnote{
On leave of absence from the Institute for High Energy Physics,
142284, Protvino, Moscow region, Russia; e-mail: saveliev@mx.ihep.su}
\\
{\footnotesize Laboratoire de Physique Th\'eorique de
l'\'Ecole Normale Sup\'erieure\footnote{Unit\'e Propre du
Centre National de la Recherche Scientifique,
associ\'ee \`a l'\'Ecole Normale Sup\'erieure et \`a l'Universit\'e
de Paris-Sud},\\
24 rue Lhomond, 75231 Paris C\'EDEX 05, ~France\footnote{
E-mail: saveliev@physique.ens.fr}}

\medskip
\end{center}

\vfill
\begin{abstract}
\baselineskip .4 true cm
\noindent
Some special solutions to the multidimensional Lam\'e and Bourlet type 
equations are constructed in an explicit form. 
\end{abstract}
\vfill
\end{titlepage}
\baselineskip .5 true cm

\setcounter{footnote}{1}

\section{Introduction}

The classical differential geometry serves as an injector of many
equations integrable in this or that sense. Among them, the Lam\'e
and Bourlet equations play especially remarkable role. These
equations arise, in particular, in the following way.

Let $(U, z_1, \ldots, z_n)$ be a chart on a Riemannian manifold
$M$, such that the metric tensor $g$ has on $U$ the form
\begin{equation}
g = \sum_{i=1}^n \beta_i^2(z) \, dz_i \otimes dz_i. \label{1.4}
\end{equation}
In such a situation the metric tensor $g$ is said to be {\it
diagonal} with respect to the coordinates $z_i$.  The functions
$\beta_i$ are called the {\it Lam\'e coefficients}. Define the so
called {\it rotation coefficients}
\begin{equation}
\gamma_{ij} = \frac{1}{\beta_i} \partial_i \beta_j, \qquad i \ne j.
\label{1.3} 
\end{equation}
Here and henceforth we denote $\partial_i = \partial/\partial z_i$.
It can be shown that the Riemannian submanifold $U$ of $M$ is flat if
and only if the rotation coefficients $\gamma_{ij}$ satisfy the
following system of partial differential equations
\begin{eqnarray}
&&\partial_i \gamma_{jk} = \gamma_{ji} \gamma_{ik}, \quad i \ne j \ne
k, \label{1.1} \\
&&\partial_i \gamma_{ij} + \partial_j \gamma_{ji} + \sum_{k \neq i,j}
\gamma_{ki} \gamma_{kj} = 0, \quad i \neq j, \label{1.2}
\end{eqnarray}
where the notation $i \ne j \ne k$ means that $i$, $j$, $k$ are
distinct. Equations (\ref{1.1}), (\ref{1.2}) are called the {\it
Lam\'e equations}. 

With the so called {\it Egoroff property}, $\gamma_{ij}=\gamma_{ji}$,
equations (\ref{1.2}) are equivalent to the following ones:
\[
\left( \sum_{k=1}^n \partial/\partial_k \right) \gamma_{ij} = 0,
\qquad i \ne j.
\]
The corresponding solutions are represented in the form
$\beta_i^2=\partial_i F$ where $F$ is some function of the
coordinates $z^i$.

It can be also shown that the Riemannian submanifold $U$ is of
constant curvature \cite{KNo63} with the sectional curvature equal to
$1$ if and only if the rotation coefficients satisfy the equations
\begin{eqnarray}
&&\partial_i \gamma_{jk} = \gamma_{ji} \gamma_{ik}, \quad i \ne j \ne
k, \label{1.5} \\
&&\partial_i \gamma_{ij} + \partial_j \gamma_{ji} + \sum_{k \neq i,j}
\gamma_{ki} \gamma_{kj} + \beta_i \beta_j = 0, \quad i \neq j.
\label{1.6} 
\end{eqnarray}
We call equations (\ref{1.5}), (\ref{1.6}) and (\ref{1.3}) the {\it
Bourlet type equations}. The Bourlet equations in the precise sense
correspond to the case with $\sum_{i=1}^n \beta_i^2 = 1$, see, for
example, \cite{Bia24,Dar10}.

Sometimes it is suitable to rewrite at least a part of these
equations in a `Laplacian' type form. Impose the condition
\begin{equation}
\sum_{i=1}^n \beta_i^2 = c,
\label{1.7}
\end{equation}
where $c$ is a constant. It is convenient to allow the functions
$\beta_i$, and hence the functions $\gamma_{ij}$, to take complex
values. Therefore, we will assume that $c$ is an arbitrary complex
number.  One can easily get convinced by a direct check with account
of (\ref{1.3}) and (\ref{1.7}) that there takes place the relation
\[
\partial_i\beta_i = - \sum_{j \neq i} \gamma_{ij} \beta_j.
\]
Now, using the same calculations as those in \cite{Ami81,Sav86}, and 
introducing, as there, the operators
\[
\Delta_{(i)} = \sum_{j \ne i} \partial_j^2 -\partial_i^2, 
\]
we obtain from equations (\ref{1.6}) 
\[
\Delta_{(i)} \beta_i = \sum_{j \ne i} \beta_i [(\beta_i^{-1}
\partial_i \beta_j)^2 - (\beta_j^{-1} \partial_j \beta_i)^2] -
2\sum_{j \ne k \ne i} \beta_j \beta_k^{-2} (\partial_k \beta_i)
(\partial_k \beta_j) + \beta_i (\beta_i^2 - c).
\]

The integrability of the equations in question has been established
for quite a long time ago; the general solution is defined by
$n(n-1)/2$ functions of two variables for the Lam\'e system, and by
$n(n-1)$ functions of one variable and $n$ constants for the Bourlet
system.  However, an explicit form of the solutions for higher
dimensions remained unknown. In the beginning of eighties an interest
to these equations was revived. In particular, it was shown that for
$\sum_{i=1}^n \beta_i^2 = 1$, the completely integrable system
(\ref{1.5}), (\ref{1.6}) provides the necessary and sufficient
condition for a construction of an arbitrary local analytic immersion
of the Lobachevsky space $L_n$ in ${\Bbb R}^{2n-1}$ \cite{Ami81}, see
also \cite{TTe80}.  For $n = 2$ equations (\ref{1.5}) and (\ref{1.1})
are absent; equation (\ref{1.6}) is reduced to the Liouville and
sine-Gordon equations for $\beta_1^2 + \beta_2^2$ equals $0$ and $1$,
respectively; while (\ref{1.2}) is the wave equation. This is why for
higher dimensions the Bourlet type equations with a nonzero
constant $c$ in (\ref{1.7}), with $c = 0$, and the Lam\'e equations
sometimes are called multidimensional generalisations of the
sine-Gordon, Liouville, and wave equations, respectively, see, for
example, \cite{Ami81,TTe80,Sav86,ABT86}. In accordance with
\cite{ABT86,Zak96}, these systems can be integrated with the help of
the inverse scattering method. Moreover, in the last paper it was
shown that the problems of description of $n$-orthogonal surfaces and
classification of Hamiltonians of hydrodynamic type systems are
almost equivalent. It was also pointed out there that system
(\ref{1.5}) is a natural generalisation of the three wave system
which is a relevant object in nonlinear optics. Finally notice that
the Lam\'e equations also arise very naturally in the context of the
Cecotti-Vafa equations describing topological-antitopological fusion,
see \cite{Dub93} and references therein, and in those of the
multidimensional generalisations of the Toda type systems
\cite{RSa96}. In general, classification and description of diagonal
metrics seems to be relevant for some modern problems of supergravity
theories, in particular their elementary and solitonic supersymmetric
$p$-brane solutions, see, for example, \cite{LPSS95} and references
therein.

In the present paper we obtain in an explicit and rather simple form
some special class of the solutions to the Lam\'e equations and to
the Bourlet type equations with and without condition (\ref{1.7}).
If one does not impose condition (\ref{1.7}), then our solutions are
determined by $n$ arbitrary functions of only one variable, while
with the condition (\ref{1.7}) the obtained solutions of the Bourlet
equations are expressed as the products of the elliptic integrals of
the first kind and are determined by $2n$ arbitrary constants.  The
derivation of the solutions to both of these systems is given by
using two different methods. One is based on the geometrical
interpretation of the corresponding equations.  Another approach uses
the zero curvature representation which, for the Lam\'e equations, is
different from
\cite{Zak96}, and for the Bourlet equations is different from \cite{ABT86}.

\section{Bourlet type equations}

We begin with the description of the zero curvature representation of
the Bourlet type equations following \cite{Sav86}.

Let ${\sf M}_{ab}$ be the elements of the Lie algebra ${\frak o}(n+1,
{\Bbb R})$ of the Lie group ${\rm O}(n+1, {\Bbb R})$ defined as
\[
({\sf M}_{ab})_{cd} = \delta_{ac} \delta_{bd} - \delta_{bc} \delta_{ad},
\]
The commutation relations for these elements have the standard form
\[
[{\sf M}_{ab}, {\sf M}_{cd}] = \delta_{ad} {\sf M}_{bc} + \delta_{bc}
{\sf M}_{ad} - \delta_{ac} {\sf M}_{bd} - \delta_{bd} {\sf M}_{ac},
\]
and any element ${\sf X}$ of ${\frak g}$ can be represented as
\[
{\sf X} = \sum_{a,b =1}^{n+1} x_{ab} {\sf M}_{ab}.
\]
Such a representation is unique if we suppose that $x_{ab} = -
x_{ba}$.

In what follows we assume that the indices $a, b, c, \ldots$ run from
$1$ to $n+1$, while the indices $i, j, k, \ldots $ run from $1$ to
$n$. Let $(U, z_1, \ldots, z_n)$ be a chart on some smooth manifold
$M$.  Consider the connection $\omega = \sum_{i = 1}^n \omega_i dz^i$
on the trivial principal fibre bundle $U \times {\rm O}(n+1, {\Bbb
R})$ with the components given by
\begin{equation}
\omega_i = \sum_{k=1}^n \gamma_{ki} {\sf M}_{ik} + \beta_i {\sf
M}_{i, n+1}. \label{2.2} 
\end{equation}
One can get convinced that the Bourlet type equations (\ref{1.3}),
(\ref{1.5}) and (\ref{1.6}) are equivalent to the zero curvature
condition for the connection $\omega$, which, in terms of the
connection components, has the form
\begin{equation}
\partial_i \omega_j- \partial_j \omega_i + [\omega_i, \omega_j] =
0. \label{2.3}
\end{equation}

Identify the Lie group ${\rm O}(n, {\Bbb R})$ with the Lie subgroup
of ${\rm O}(n+1, {\Bbb R})$ formed by the matrices ${\sf A} \in {\rm
O}(n+1, {\Bbb R})$, such that
\[
{\sf A}_{i, n+1} = 0, \qquad {\sf A}_{n+1, j} = 0, \qquad {\sf
A}_{n+1, n+1} = 1. 
\]
Similarly, identify the Lie algebra ${\frak o}(n, {\Bbb R})$ with the
corresponding subalgebra of ${\frak o}(n+1, {\Bbb R})$. 

Let the connection $\omega$ with the components of form (\ref{2.2})
satisfies the zero curvature condition (\ref{2.3}). Suppose that $U$ is
simply connected, then there exists a mapping $\varphi$ from $U$ to
${\rm O}(n+1, {\Bbb R})$, such that
\[
\omega_i = \varphi^{-1} \partial_i \varphi.
\]
Parametrise $\varphi$ in the following way
\begin{equation}
\varphi = \xi \chi, \label{2.10}
\end{equation}
where $\chi$ is a mapping from $U$ to ${\rm O}(n, {\Bbb R})$ and the
mapping $\xi$ has the form 
\begin{equation}
\xi = e^{\psi_1 {\sf M}_{12}} e^{\psi_2 {\sf M}_{23}} \cdots
e^{\psi_{n-1} {\sf M}_{n-1,n}} e^{\psi_n {\sf M}_{n,n+1}}.
\label{2.4}
\end{equation}
Here $\psi_i$ are some functions on $U$ having the meaning of the
generalised Euler angles \cite{Vil69}. For the connection components
$\omega_i$ one obtains the expression
\[
\omega_i = \chi^{-1} (\xi^{-1} \partial_i \xi) \chi + \chi^{-1}
\partial_i \chi.
\]
Relation (\ref{2.4}) gives
\[
\xi^{-1} \partial_i \xi = \sum_{j=1}^{n-1} \partial_i \psi_j \sum_{k
= j+1}^n \mu_{jk} (\psi) \, {\sf M}_{jk} + \sum_{j=1}^n \partial_i \psi_j
\, \nu_j (\psi) \, {\sf M}_{j, n+1},
\]
where
\begin{eqnarray}
&&\mu_{j-1, j}(\psi) = \cos \psi_j, \quad 1 < j \le n, \label{2.30} \\ 
&&\mu_{jk}(\psi) = \left( \prod_{l=j+1}^{k-1} \sin \psi_l \right)
\cos \psi_k, \quad 1 < j+1 < k \le n, \label{2.31} \\
&&\nu_j (\psi) = \prod_{l =j+1}^n \sin \psi_l, \quad 1  \le j < n, \qquad
\nu_n(\psi) = 1. \label{2.32}
\end{eqnarray}
Now, using the evident equalities
\begin{equation}
\chi^{-1} \partial_i \chi = \frac{1}{2} \sum_{j,k,l =1}^n \chi_{lj}
\partial_i \chi_{lk} {\sf M}_{jk}, \qquad \chi^{-1} {\sf M}_{i, n+1}
\chi = \sum_{j=1}^n \chi_{ij} {\sf M}_{j, n+1},\label{2.14} 
\end{equation}
one comes to the expressions
\begin{eqnarray}
\omega_i &=& \frac{1}{2} \sum_{j,k,l=1}^n \chi_{lj} \, \partial_i
\chi_{lk} \, {\sf M}_{jk} \nonumber \\
&& \hskip 2em {} + \sum_{j,k=1}^n \sum_{l=1}^{n-1} \partial_i \psi_l
\sum_{m = l+1}^n \mu_{lm}(\psi) \, \chi_{lj} \, \chi_{mk} \, {\sf
M}_{jk} + \sum_{j,l=1}^n \partial_i \psi_l \, \nu_l(\psi) \,
\chi_{lj} \, {\sf M}_{j, n+1}. \hskip 2em \label{2.5} 
\end{eqnarray}
Comparing (\ref{2.5}) and (\ref{2.2}), we have, in particular,
\begin{equation}
\sum_{l=1}^n \partial_i \psi_l \, \nu_l(\psi) \, \chi_{lj} =
\beta_i \, \delta_{ij}.
\label{2.6} 
\end{equation}
Note that the geometrical meaning of the functions $\beta_i$ do not
allow them to take zero value. Therefore, from (\ref{2.6}) it follows
that for any point $p \in U$ we have
\begin{equation}
\det (\partial_i \psi_j(p)) \ne 0, \qquad  \nu_i(\psi(p)) \ne 0.
\label{2.8} 
\end{equation}
Since the matrix $(\chi_{ij})$ is orthogonal, one easily obtains
\[
\chi_{ij} = \frac{1}{\beta_j} \partial_j \psi_i \, \nu_i(\psi),
\]
and, using again the orthogonality of $(\chi_{ij})$, one sees that 
\begin{equation}
\beta_i^2 = \sum_{l=1}^n (\partial_i \psi_l \, \nu_l(\psi))^2.
\label{2.29} 
\end{equation}
Therefore, we have
\begin{equation}
\chi_{ij} = \frac{\partial_j \psi_i \, \nu_i(\psi)}{\sqrt{\sum_{l=1}^n
(\partial_j \psi_l \, \nu_l(\psi))^2}}. \label{2.9}  
\end{equation}
Thus, the matrix $(\chi_{ij})$, and hence the mapping $\chi$, is
completely determined by the functions $\psi_i$, and its
orthogonality is equivalently realised by the relation
\begin{equation}
\sum_{l=1}^n \partial_i \psi_l \, \nu_l^2(\psi) \, \partial_j \psi_l
= 0, \qquad i \ne j. 
\label{2.7}
\end{equation}

Suppose now that a set of functions $\psi_i$ satisfies relations
(\ref{2.8}) and (\ref{2.7}). Consider the mapping $\varphi$ defined
by (\ref{2.10}) with the mapping $\xi$ having form (\ref{2.4}) and
the mapping $\chi$ defined by (\ref{2.9}). Show that the mapping
$\varphi$ generates the connection with the components of form
(\ref{2.2}).  First of all, with $\beta_i$ of form (\ref{2.29}) we
can get convinced that in the case under consideration relation
(\ref{2.6}) is valid. Taking into account (\ref{2.7}), one can write
the relation
\[
\sum_{l=1}^n \partial_j \psi_l \, \nu_l^2 (\psi) \, \partial_k \psi_l
= \beta^2_j \, \delta_{jk}, 
\] 
whose differentiation with respect to $z_i$ gives
\begin{eqnarray*}
&&\sum_{l=1}^n \partial_j \psi_l \, \nu_l^2(\psi) \, \partial_i
\partial_k \psi_l \\
&&\hskip 2em {} = - \sum_{l=1}^n \partial_i \partial_j \psi_l \,
\nu_l^2(\psi) \, \partial_k \psi_l - 2 \sum_{l=1}^n \partial_j
\psi_l \, \nu_l(\psi) \, \partial_i \nu_l(\psi) \, \partial_k \psi_l
+ 2 \beta_j \partial_i \beta_j \, \delta_{jk}.
\end{eqnarray*}
Since the left hand side of this equality is symmetric with respect
to the transposition of the indices $i$ and $k$, its right hand side
must also be symmetric with respect to this transposition, and,
therefore, we have 
\begin{eqnarray*}
&&\sum_{l=1}^n \partial_j \psi_l \, \nu_l^2(\psi) \, \partial_i
\partial_k \psi_l \\
&& \hskip 2em {} = - \sum_{l=1}^n \partial_k \partial_j \psi_l \,
\nu_l^2(\psi) \, \partial_i \psi_l - 2 \sum_{l=1}^n \partial_j
\psi_l \, \nu_l(\psi) \, \partial_k \nu_l(\psi) \, \partial_i \psi_l
+ 2 \beta_j 
\partial_k \beta_j \, \delta_{ij}.
\end{eqnarray*}
Using this equality, it is not difficult to show that
\begin{eqnarray}
\sum_{l=1}^n \chi_{lj} \, \partial_i \chi_{lk} &=& \gamma_{kj} \,
\delta_{ij} - \gamma_{jk} \, \delta_{ik} \nonumber \\
&-& \frac{1}{\beta_j \beta_k} \sum_{l=1}^n [\partial_j \psi_l \,
\nu_l(\psi) \, \partial_k \nu_l(\psi) \, \partial_i \psi_l -
\partial_k \psi_l \, \nu_l(\psi) \,
\partial_j \nu_l(\psi) \, \partial_i \psi_l], \label{2.11} 
\end{eqnarray}
where the functions $\gamma_{ij}$ are defined by (\ref{1.3}).

Using the concrete form of the functions $\mu_{ij}(\psi)$ and
$\nu_i(\psi)$, we can get convinced in the validity of the equalities
\[
\frac{\partial \nu_j(\psi)}{\partial \psi_i} = 0, \quad 1 \le i
\le j, \qquad \mu_{ij}(\psi) = \frac{1}{\nu_j(\psi)} \frac{\partial
\nu_i(\psi)}{\partial \psi_j},\]
which allow to show that 
\begin{equation}
\sum_{l=1}^{n-1} \partial_i \psi_l \sum_{m = l+1}^n
\mu_{lm}(\psi) \, \chi_{lj} \, \chi_{mk} = \frac{1}{\beta_j \beta_k}
\sum_{l=1}^n \partial_j \psi_l \, \nu_l(\psi) \, \partial_k \nu_l(\psi) \,
\partial_i \psi_l. \label{2.23}
\end{equation}

Substituting (\ref{2.11}), (\ref{2.23}) and (\ref{2.6}) into
(\ref{2.5}), we come to expression (\ref{2.2}). Thus, any set of
functions $\psi_i$ satisfying (\ref{2.8}) and (\ref{2.7}) allows to
construct a connection of form (\ref{2.2}) satisfying the zero
curvature condition (\ref{2.3}) which is equivalent to the Bourlet
type equations. Therefore, the general solution to the
Bourlet type equations is described by (\ref{2.29}) where the
functions $\psi_i$ satisfy (\ref{2.8}) and (\ref{2.7}). In the
simplest case we can satisfy (\ref{2.7}) assuming that
\begin{equation}
\partial_i \psi_j = 0, \qquad i \ne j; \label{2.33}
\end{equation}
in other words, each function $\psi_i$ depends on the corresponding
coordinate $z_i$ only. In this case we obtain the following
expressions for the functions $\beta_i$:
\begin{equation}
\beta_i = \partial_i \psi_i \prod_{j=i+1}^n \sin \psi_j, \quad
1 \leq i < n, \qquad \beta_n = \partial_n \psi_n. \label{2.13}
\end{equation}
The corresponding expressions for the functions $\gamma_{ij}$ can be
easily found and we do not give here their explicit form.

There is a transparent geometrical interpretation of the results
obtained above.  Recall that solutions of the Bourlet type equations
are associated with diagonal metrics in Riemannian spaces of constant
curvature. Namely, let $(U, z_1, \ldots, z_n)$ be a chart on a
manifold $M$, and we have a solution of the Bourlet type equations.
Supply the open submanifold $U$ with metric (\ref{1.4}); then $(U,
g)$ is a Riemannian manifold of constant curvature with the sectional
curvature equal to $1$. From the other hand, let $(M, g)$ be a
Riemannian manifold of constant curvature with the sectional curvature
equal to $1$, and $(U, z_1, \ldots, z_n)$ be such a chart on $M$ that
the metric $g$ has on $U$ form (\ref{1.4}). Then the Lam\'e
and the corresponding rotation coefficients satisfy the
Bourlet type equations.

The simplest example of a manifold of constant curvature is a unit
$n$-dimensional sphere $S^n$ in ${\Bbb R}^{n+1}$ with the metric
induced by the standard metric on ${\Bbb R}^{n+1}$. Denote the
spherical coordinates in $S^n$ by $z_1, \ldots, z_n$. The explicit
expression for the metric on $S^n$ has the form  
\[
g = \sum_{l=1}^{n-1} \left( \prod_{m=l+1}^n \sin^2 z_m \right)
dz_l \otimes dz_l + dz_n \otimes dz_n.
\]
So we have a diagonal metric. Note that it can be written in the form
\begin{equation}
g = \sum_{l=1}^n \nu^2_l(z) \, dz_l \otimes dz_l, \label{2.34}
\end{equation}
where the functions $\nu_i$ are given by (\ref{2.32}). Let $\psi$ be
a diffeomorphism from $S^n$ to $S^n$. It is clear that $(S^n, \psi^*
g)$ is also a Riemannian manifold of constant curvature with the
sectional curvature equal to $1$. Denoting $\psi^* z_i = \psi_i$, one
gets
\[
\psi^* g = \sum_{j,k,l=1}^n \partial_j \psi_l \, \nu^2_l(\psi) \,
\partial_k \psi_l \, dz_j \otimes dz_k.
\]
Therefore, the metric $\psi^* g$ is diagonal with respect to the
coordinates $z_i$ if and only if the functions $\psi_i$ satisfy
relations (\ref{2.7}). In particular, if the functions $\psi_i$
satisfy relations (\ref{2.33}) we obtain the diagonal metrics with the
Lam\'e coefficients given by (\ref{2.13}).

In general, starting from some fixed diagonal metric in the space of
constant curvature with the unit sectional curvature, one gets the
family of explicit solutions to the Bourlet type equations
parametrised by a set of $n$ functions each depending only on one
variable.  In terms of equations (\ref{1.3}), (\ref{1.5}) and
(\ref{1.6}) themselves, we formulate this observation as follows. Let
the functions $\beta_i$, $\gamma_{ij}$ satisfy the Bourlet type
equations; then for any set of functions $\psi_i$, such that
\[
\partial_i \psi_j = 0, \qquad i \ne j,
\]
the functions
\begin{equation}
\beta'_i(z) = \beta_i(\psi(z)) \, \partial_i \psi_i(z), \qquad
\gamma'_{ij} (z) = \gamma_{ij} (\psi(z)) \, \partial_j \psi_j(z_j),
\label{2.35} 
\end{equation}
where $\psi(z)$ stands for the set $\psi_1(z), \ldots \psi_n(z)$,
also satisfy the Bourlet type equations. 

Note that our considerations can be easily generalised to the case of
complex metrics. In this case the zero curvature representation of
the Bourlet type equations should be based on the Lie group ${\rm
O}(n+1, {\Bbb C})$.

Return to the consideration of solutions (\ref{2.13}) to the Bourlet 
type equations.  If one imposes condition
(\ref{1.7}) where $c$ is an arbitrary zero or nonzero constant, then
the arbitrary functions $\psi_i (z_i)$ satisfy the equation
\[
\sum_{l = 1}^n \left( \prod_{m=l+1}^n \sin^2 \psi_m \right) (\partial_l \psi_l)^2
= c,
\]
thereof for some constants $c_i$, $i = 0, \ldots, n$, such that $c_0
= 0$ and $c_n = c$, one gets
\begin{equation}
\partial_i \psi_i= \sqrt{c_i - c_{i-1} \sin^2 \psi_i}. \label{el1}
\end{equation}
Hence, solution (\ref{2.13}) takes the form
\begin{equation}
\beta_i =  \sqrt{c_i - c_{i-1} \sin^2 \psi_i} \prod_{j=i+1}^n \sin
\psi_j, \label{a44}
\end{equation}
where the functions $\psi_i$ are determined by the ordinary
differential equations (\ref{el1}). Suppose that all constants $c_i$,
$i = 1,\ldots,n$, are different  from zero. With appropriate
conditions on the constants $c_i$, in accordance with (\ref{el1}) one
has
\[
z_i + d_i = \int^{\psi_i}_0 \frac{d\psi_i}{\sqrt{c_i-c_{i-1} \sin^2
\psi_i}}, 
\]
where $d_i$ are arbitrary constants. Therefore,
\[
\sqrt{c_i}(z_i + d_i) = F \left( \psi_i, \,
\sqrt{\frac{c_{i-1}}{c_i}}\, \right),
\]
where $F(\phi, k)$ is the elliptic integral of the first kind,
\[
F(\phi, k)=\int^{\phi}_0 \frac{d\phi}{\sqrt{1 - k^2 \sin^2 \phi}}.
\]
Thus, using Jacobi elliptic functions, we can write
\[
\sin \psi_i(z_i) = \sn \left( \sqrt{c_i} (z_i+d_i), \,
\sqrt{\frac{c_{i-1}}{c_i}} \, \right), \qquad \cos \psi_i(z_i) = \cn
\left( \sqrt{c_i} (z_i+d_i), \, \sqrt{\frac{c_{i-1}}{c_i}} \,
\right). 
\]
Now, with the evident relation
\[
\partial_i \psi_i(z_i) = \frac{\partial_i \sin \psi_i(z_i)}{\cos
\psi_i(z_i)} = \sqrt{c_i} \dn \left( \sqrt{c_i} (z_i+d_i), \,
\sqrt{\frac{c_{i-1}}{c_i}} \, \right),
\]
we write our solution as the product of elliptic functions,
\begin{equation}
\beta_i(z) = \sqrt{c_i} \dn \left( \sqrt{c_i} (z_i+d_i), \,
\sqrt{\frac{c_{i-1}}{c_i}} \, \right) \prod_{j=i+1}^n \sn \left( \sqrt{c_j}
(z_j+d_j), \, \sqrt{\frac{c_{j-1}}{c_j}} \, \right).\label{rez}
\end{equation}
The case when some of the constants $c_i$ are equal to zero can be
analysed in a similar way.
Note that, taking into account the relations
\[
\sn (u, 1) = \tanh u, \quad \dn(u, 1) = \frac{1}{\cosh u}, \quad \sn
(u, 0) = \sin u, \quad \dn (u, 0) = 1, 
\]
with an appropriate choice of the constants $c_i$, we can reduce some
of the elliptic functions entering the obtained solution to the
trigonometric or hyperbolic ones.

It is clear from the solution in form (\ref{a44}) or (\ref{rez}),
that it does not depend on the variable $z_1$ at all, since among
the functions $\beta_i$, only $\beta_1$ depends on $\psi_1$ and only
as $\partial_1\psi_1$, while $\psi_1= c_1 z_1+d_1$.

In the simplest case $n=2$ and $c_2=1$ with the parametrisation
$\beta_1 = \cos(u/2)$, $\beta_2 = \sin(u/2)$, system (\ref{1.3}),
(\ref{1.5}) and (\ref{1.6}) is reduced to the sine-Gordon equation
\[
\partial_1^2 u - \partial_2^2 u + \sin u = 0,
\]
and one gets the evident solution $\sin(u/2) = \dn (z_2 + d_2, \sqrt{c_1})$.

\section{Lam\'e equations}

The zero curvature representation of the Lam\'e equations is based on
the Lie group $G$ of rigid motions of the affine space ${\Bbb R}^n$.
This Lie group is isomorphic to the semidirect product of the Lie
groups ${\rm O}(n, {\Bbb R})$ and ${\Bbb R}^n$, where the linear
space ${\Bbb R}^n$ is considered as a Lie group with respect to the
addition operation.  The standard basis of the Lie algebra ${\frak
g}$ of the Lie group $G$ consists of the elements ${\sf M}_{ij}$ and
${\sf P}_i$ which satisfy the commutation relations
\begin{eqnarray*}
&[{\sf M}_{ij}, {\sf M}_{kl}] = \delta_{il} {\sf M}_{jk} +
\delta_{jk} {\sf M}_{il} - \delta_{ik} {\sf M}_{jl} - \delta_{jl}
{\sf M}_{ik},& \\ 
&[{\sf M}_{ij}, {\sf P}_k] = \delta_{jk} {\sf P}_i - \delta_{ik} {\sf
P}_j, \qquad [{\sf P}_i, {\sf P}_j] = 0.
\end{eqnarray*}

Let $(U, z_1, \ldots, z_n)$ be a chart on the manifold $M$.  Consider
the connection $\omega = \sum_{i = 1}^n \omega_i dz^i$ on the trivial
principal fibre bundle $U \times G$ with the components given by
\begin{equation}
\omega_i = \sum_{k=1}^n \gamma_{ki} {\sf M}_{ik} + \beta_i {\sf P}_i.
\label{4.2} 
\end{equation}
It can be easily verified that equations (\ref{1.3})--(\ref{1.2}) are
equivalent to the zero curvature condition for the connection
$\omega$. It is well known that the Lie algebra ${\frak g}$ can be
obtained from the Lie algebra ${\frak o}(n+1, {\Bbb R})$ by an
appropriate In\"on\"u--Wigner contraction. Unfortunately, this fact
does not give us a direct procedure for obtaining solutions of the
Lam\'e equations from solutions of the Bourlet type equations.
Therefore, we will consider the procedure for obtaining solutions of
the Lam\'e equations independently.

Let the connection $\omega$ with the components of form (\ref{4.2})
satisfies the zero curvature condition. Restricting to the case of
simply connected $U$, write for the connection components $\omega_i$
the representation
\[
\omega_i = \varphi^{-1} \partial_i \varphi,
\]
where $\varphi$ is some mapping from $U$ to $G$. Parametrise
$\varphi$ in the following way:
\begin{equation}
\varphi = \xi \chi \label{4.10}
\end{equation}
where $\chi$ is a mapping from $U$ to ${\rm O}(n, {\Bbb R})$,
and the mapping $\xi$ has the form
\begin{equation}
\xi = e^{\psi_1 {\sf P}_1} e^{\psi_2 {\sf P}_2} \cdots e^{\psi_{n-1}
{\sf P}_{n-1}} e^{\psi_n {\sf P}_n}. \label{4.4} 
\end{equation}
For the connection components $\omega_i$ one obtains the expression
\begin{equation}
\omega_i = \frac{1}{2} \sum_{j,k,l=1}^n \chi_{lj} \partial_i
\chi_{lk} {\sf M}_{jk} + \sum_{j,l=1}^n \partial_i \psi_l \chi_{lj}
{\sf P}_j. 
\label{4.5} 
\end{equation}
{}From the comparison of (\ref{4.5}) and (\ref{4.2}) we see that
\begin{equation}
\chi_{ij} = \frac{\partial_j \psi_i}{\sqrt{\sum_{l=1}^n (\partial_j
\psi_l)^2}}, \label{4.9}  
\end{equation}
and the functions $\psi_i$ satisfy the relation
\begin{equation}
\sum_{l=1}^n \partial_i \psi_l \, \partial_j \psi_l = 0, \qquad i \ne j.
\label{4.7}
\end{equation}
The functions $\beta_i$ are connected with the functions $\psi_i$ by
the formula
\begin{equation}
\beta_i^2 = \sum_{l=1}^n (\partial_i \psi_l)^2, \label{4.6}
\end{equation}
and from the geometrical meaning of $\beta_i$ it follows that
\begin{equation}
\det (\partial_i \psi_j(a)) \ne 0. \label{4.8}
\end{equation}

Suppose now that a set of functions $\psi_i$ satisfies relations
(\ref{4.7}) and (\ref{4.8}). Consider the mapping $\varphi$ defined
by (\ref{4.10}) with the mapping $\xi$ having form (\ref{4.4}) and
the mapping $\chi$ defined by (\ref{4.9}). It can be shown that the
mapping $\varphi$ generates the connection with the components of
form (\ref{4.2}). Here the functions $\beta_i$ are defined from
(\ref{4.6}), and the functions $\gamma_{ij}$ are given by
(\ref{1.3}). Thus, any set of functions $\psi_i$ satisfying
(\ref{4.8}) and (\ref{4.7}) allows to construct a connection of form
(\ref{4.2}) satisfying the zero curvature condition which is
equivalent to the Lam\'e equations, and in such a way we obtain the
general solution.

Assuming that the functions $\psi_i$ satisfy (\ref{2.33}), we have
\begin{equation}
\beta_i = \partial_i \psi_i. \label{4.11}
\end{equation}
It is clear that in this case $\gamma_{ij} = 0$. So one ends up with a
trivial solution of the Lam\'e equations. To get nontrivial solutions
one should consider different parametrisations of the mapping
$\varphi$.  For example, let us represent the mapping $\varphi$ in
form (\ref{4.10}) where the mapping $\chi$ again takes values in
${\rm O}(n, {\Bbb R})$, while the mapping $\xi$ has the form
\[
\xi = e^{\psi_1 {\sf M}_{12}} e^{\psi_2 {\sf M}_{23}} \cdots e^{\psi_{n-1}
{\sf M}_{n-1, n}} e^{\psi_n {\sf P}_n}. 
\]
With such a parametrisation of $\xi$, one gets
\begin{eqnarray*}
\omega_i &=& \frac{1}{2} \sum_{j,k,l=1}^n \chi_{lj} \, \partial_i
\chi_{lk} \, {\sf M}_{jk} \\
&&\hskip 2.em {} + \sum_{j,k=1}^n \sum_{l=1}^{n-1} \partial_i \psi_l
\sum_{m = l+1}^n \mu_{lm}(\psi) \, \chi_{lj} \, \chi_{mk} \, {\sf
M}_{jk} + \sum_{j,l=1}^n \partial_i \psi_l \, \nu_l(\psi) \,
\chi_{lj} \, {\sf P}_j, \hskip 2.em 
\end{eqnarray*}
where
\begin{eqnarray}
&&\mu_{j-1, j}(\psi) = \cos \psi_j, \quad 1 < j < n; \qquad
\mu_{n-1, n}(\psi) = 1; \label{4.30} \\
&&\mu_{jk}(\psi) = \left( \prod_{l=j+1}^{k-1} \sin \psi_l \right)
\cos \psi_k, \quad 1 < j+1 < k < n; \label{4.31} \\
&&\mu_{jn}(\psi) = \prod_{l=j+1}^{n-1} \sin \psi_l, \quad 1 < j+1 <
n; \label{4.32} \\ 
&&\displaystyle \nu_j (\psi) = \left( \prod_{k =j+1}^{n-1} \sin
\psi_k \right) \psi_n, \; 1 \le j < n-1; \quad \nu_{n-1}(\psi) =
\psi_n; \quad \nu_n(\psi) = 1. \label{4.33}
\end{eqnarray}
Using these relations we come to the following description of the
general solution to the Lam\'e equations. Let functions $\psi_i$
satisfy the relations
\[
\sum_{l=1}^n \partial_i \psi_l \, \nu_l^2(\psi) \, \partial_j \psi_l
= 0, \qquad i \ne j,
\]
and for any point $p \in U$ one has
\[
\det (\partial_i \psi_j(p)) \ne 0, \qquad  \nu_i(\psi(p)) \ne 0.
\]
Then the functions $\beta_i$ determined from the equality
\[
\beta_i^2 = \sum_{l=1}^n (\partial_i \psi_l \, \nu_l(\psi))^2,
\]
and the corresponding functions $\gamma_{ij}$ defined by (\ref{1.3})
give the general solution of the Lam\'e equations.
If the functions $\psi_i$ satisfy (\ref{2.33}), we get the following
expressions for the functions $\beta_i$
\begin{equation}
\displaystyle \beta_i = \partial_i \psi_i \left( \prod_{j=i+1}^{n-1}
\sin \psi_j \right) \psi_n, \; 1 \leq i < n-1,\quad
\beta_{n-1} = \partial_{n-1} \psi_{n-1} \, \psi_n, \quad \beta_n =
\partial_n \psi_n. \label{4.34}
\end{equation}

The geometrical interpretation of the obtained solutions to the
Lam\'e equations is similar to one given in the previous section.
Recall that solutions of the Lam\'e equations are associated with
flat diagonal metrics in flat Riemannian spaces. The simplest case
here is the standard metric in ${\Bbb R}^n$,
\[
g = \sum_{l=1}^n dz_l \otimes dz_l.
\]
Applying a diffeomorphism $\psi$ one gets the metric
\[
\psi^* g = \sum_{j,k,l=1}^n \partial_j \psi_l \partial_k \psi_l \, dz_j
\otimes dz_k,
\] 
which is diagonal if and only if the functions $\psi_i$ satisfy
(\ref{4.7}). Here the functions $\psi_i$ which obey (\ref{2.33}) give
the Lam\'e coefficients described by (\ref{4.11}).

A more nontrivial example is provided by the metric arising after the
transition to the spherical coordinates in ${\Bbb R}^n$. Denoting the
standard coordinates on ${\Bbb R}^n$ by $x_i$ and the spherical
coordinates by $r$ and $\theta_1, \ldots \theta_{n-1}$, one has
\[
x_1 = r \prod_{k=1}^{n-1} \sin \theta_k,\quad x_i = r \cos \theta_{i-1} 
\prod_{k = i}^{n-1} \sin \theta_k,\; 1<i<n,\quad x_n = r \cos
\theta_{n-1}. 
\]
For the metric we obtain the expression
\[
g = r^2 \left[ \sum_{l=1}^{n-2} \left( \prod_{m=l+1}^{n-1}
\sin^2 \theta_m \right) d\theta_l \otimes d\theta_l + d\theta_{n-1}
\otimes d\theta_{n-1} \right] + dr \otimes dr.
\]
Denoting $z_i = \theta_i$, $i=1,\ldots,n-1$, and $z_n = r$, we rewrite
this relation as
\[
g = z_n^2 \left[ \sum_{l=1}^{n-2} \left( \prod_{m=l+1}^{n-1}
\sin^2 z_m \right) dz_l \otimes dz_l + dz_{n-1} \otimes dz_{n-1}
\right] + dz_n \otimes dz_n. 
\]
Therefore, the metric $g$ has form (\ref{2.34}) with the functions
$\nu_i$ given by (\ref{4.33}). A diffeomorphism $\psi$ with the
functions $\psi_i = \psi^* z_i$ satisfying (\ref{2.33}) gives the
metric with the Lam\'e coefficients (\ref{4.34}).

In conclusion note that relations (\ref{2.35}) describe the
symmetry transformations not only of the Bourlet type equations, but
also of the Lam\'e equations, and actually the existence of such
transformations allows us to construct the solutions of the equations
under consideration parametrised by $n$ arbitrary functions each
depending on one variable.

\section*{Acknowledgements}

The authors are grateful to B.~A.~Dubrovin, J.-L.~Gervais,
V.~A.~Kazakov, Yu.~I.~Manin and Yu.~G.~Stro\-ga\-nov for the
discussions. We also thank V.~E.~Zakharov for acquainting us with his
recent results prior to publication.  One of the authors (M.~V.~S.)
wishes to acknowledge the warm hospitality and creative scientific
atmosphere of the Laboratoire de Physique Th\'eorique de l'\'Ecole
Normale Sup\'erieure de Paris. This work was supported in part by the
Russian Foundation for Basic Research under grant no. 95--01--00125a.

\end{document}